\documentclass[conference]{IEEEtran}
\ifCLASSINFOpdf
  \usepackage[pdftex]{graphicx}
   \graphicspath{{./}}
\else
   \usepackage[dvips]{graphicx}
   \graphicspath{{./}}
\fi

\usepackage{amsmath,amsfonts,bm,graphicx,hyperref}
\newcommand{\llangle}{\langle\!\langle}
\newcommand{\rrangle}{\rangle\!\rangle}

\begin{document}
\title{Noise and Counting Statistics of a\\ Single Electron Emitter: Theory}

\author{\IEEEauthorblockN{Mathias Albert,
Christian Flindt and
Markus B\"uttiker}
\IEEEauthorblockA{D\'epartement de Physique Th\'eorique,
Universit\'e de Gen\`eve, CH-1211 Gen\`eve, Switzerland}}

\IEEEspecialpapernotice{(Invited Paper)}

\maketitle

\begin{abstract}
We review the latest progress in understanding the fundamental noise properties of a coherent single electron emitter known as the mesoscopic capacitor. The system consists of a sub-micron cavity connected to a two-dimensional electron gas via a quantum point contact. When subject to periodic gate voltage modulations, the mesoscopic capacitor absorbs and re-emits single electrons at giga-hertz frequencies as it has been demonstrated experimentally. Recent high-frequency noise measurements have moreover allowed for a precise characterization of the device in different operating regimes. Here we discuss a simple model of the basic charge transfer processes in the mesoscopic capacitor and show how the model is capable of fully reproducing the measured high-frequency noise spectrum. We extend our analysis to the counting statistics of emitted electrons during a large number of periods which we use to discuss the accuracy of the mesoscopic capacitor as a single electron source. Finally, we consider possible applications of the mesoscopic capacitor in future experiments and identify novel pathways for further theoretical research.
\end{abstract}

\section{Introduction}

\begin{figure}
  \begin{center}
    \includegraphics[width=0.95\linewidth]{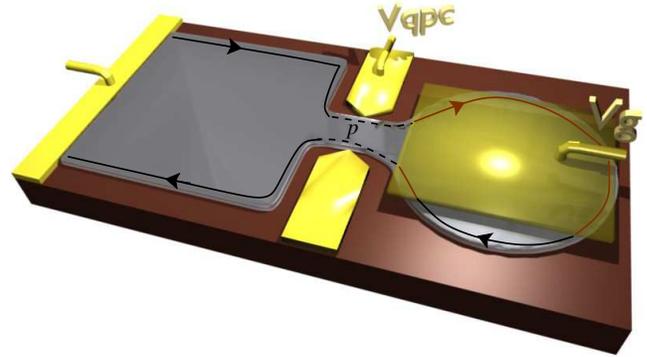}
    \caption{Mesoscopic capacitor. The mesoscopic capacitor consists of a nano-scale cavity coupled via a quantum point contact to a single state running along the edge of a two-dimensional electron gas. The periodically modulated voltage $V_g(t)$ causes absorbtion and emission of single electrons to and from the edge state.}
    \label{fig:system}
  \end{center}
\end{figure}

Controllable single electron sources are expected to form a fundamental building block in future electronic circuits whose functionality is based solely on a single or a few electrons rather than operating with macroscopically large currents as in conventional electronics.  The mesoscopic capacitor constitutes one archetype of such a single electron emitter. The system consists of a nano-scale cavity that is coupled via a quantum point contact to an edge state of a two-dimensional electron gas, Fig.~\ref{fig:system}.\footnote{We thank S.\ E.\ Nigg for providing us with the figure.} Single electrons can tunnel between the edge state and the mesoscopic capacitor with a tunneling amplitude that is experimentally controllable. The mesoscopic capacitor is important because of its promising applications in future electronics, but it is also an elementary quantum device which is interesting from a purely scientific point of view.

The mesoscopic capacitor has been the subject of intensive experimental and theoretical research. Originally, the mesoscopic capacitor was proposed by B\"uttiker \emph{et al.} who showed that the relaxation resistance is quantized in units of $h/2e^2$ \cite{But93}, independently of the microscopic details \cite{Nig06,Mor10,Ham10}. This prediction was confirmed in experiments by Gabelli \emph{et al.} \cite{Gab06}. In a subsequent experiment, F\'{e}ve \emph{et al.} additionally showed that the mesoscopic capacitor can absorb and re-emit single electrons at giga-hertz frequencies when subject to periodic gate voltage modulations \cite{Fev07}.

In this paper we review the latest progress in understanding the fundamental noise properties of the mesoscopic capacitor. Recently, Mah\'{e} \emph{et al.} measured the noise power spectrum of the mesoscopic capacitor in the giga-hertz regime \cite{Mah10}. Additionally, they proposed a simple model which reproduces the measured spectrum as well as full scattering matrix calculations in numerical simulations of the experiment \cite{Mah10}. Here, we discuss our analytic calculations of the noise spectrum and the counting statistics of emitted electrons which we use to characterize the accuracy of the mesoscopic capacitor as a single electron emitter. We consider possible applications of the mesoscopic capacitor in future experiments and identify novel pathways for further theoretical research. The present paper is partially based on our recent work reported in Ref.\ \cite{Alb10}.

The outline of the paper is as follows: In Section \ref{sec:cap} we introduce the mesoscopic capacitor. A theoretical model of the charge transport is described in Section \ref{sec:model}. Sections \ref{sec:current} and \ref{sec:spectrum} are devoted to analytic calculations of mean current and noise spectrum, respectively. The counting statistics of emitted electrons is considered in Sections \ref{sec:fcs} and \ref{sec:cumu}. In Section \ref{sec:outlook} we give an outlook on future applications and theoretical studies of the mesoscopic capacitor, before finally presenting our conclusions in Section \ref{sec:conclu}.

\section{Mesoscopic capacitor}
\label{sec:cap}

The mesoscopic capacitor is shown schematically in Fig.~\ref{fig:system}. The system consists of a nano-scale cavity connected to a two-dimensional electron gas via a quantum point contact. A strong magnetic field is applied to the sample and the system is operated in the integer quantum hall regime at filling factor $\nu=1$. Electrons close to the Fermi energy therefore propagate uni-directionally along a single state located along the edge of the sample. When an electron in the in-coming edge state arrives at the quantum point contact, it either enters the mesoscopic capacitor with probability $p$ or it is reflected to the out-going edge state with probability $1-p$. The transmission probability $p$ of the quantum point contact can be controlled experimentally using external voltage gates. Inside the mesoscopic capacitor, an electron may travel one or several rounds along the edge of the capacitor, with probability $p$ of escaping to the out-going edge state each time the electron passes the quantum point contact. With the quantum point contact completely pinched off, the mesoscopic capacitor has a discrete energy spectrum with typical level spacing $\Delta$, which is much larger than typical temperatures, and $\tau_o= h/\Delta$ is the time it takes to make a full round trip inside the capacitor.

\begin{figure}
  \begin{center}
    \includegraphics[width=0.35\textwidth]{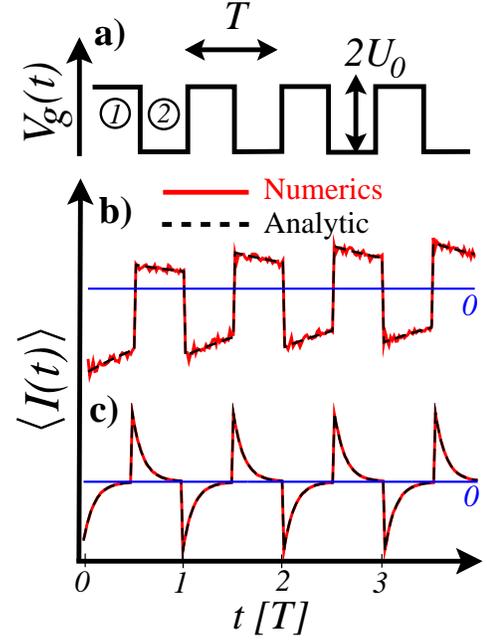}
    \caption{Periodic driving and mean current. (a) Periodic voltage applied to the capacitor. $\textcircled{\footnotesize{1}}$ denotes the absorbtion phase and $\textcircled{\footnotesize{2}}$ the emission phase. (b) Average current for $\tau/T=1$ starting with the mesoscopic capacitor being empty at $t=0$. The mean current eventually becomes periodic with period $T$. (c) Average current for  $\tau/T=0.1$. For (b) and (c), numerical data, obtained from an average over 50.000 trajectories, are compared with analytical predictions.}
    \label{fig:avcurrent}
  \end{center}
\end{figure}

With the quantum point contact being open, the electronic states of the capacitor below the Fermi level of the external electron reservoir are filled. Next, we consider the situation where a step-like gate voltage $V_g(t)$ periodically shifts the highest occupied level above and below the Fermi level of the reservoir, Fig.\ \ref{fig:avcurrent}a, corresponding to the recent experiment by Mah\'{e} \emph{et al.} \cite{Mah10}. The period of the voltage variations $T$ is much longer than $\tau_o$ and the amplitude $U_0$ is on the order of the level spacing, $2U_0e\simeq\Delta$. This causes periodic emission of coherent single electron wave packets from the capacitor to the outgoing edge state, followed by refilling from the incoming edge state as it has been demonstrated experimentally \cite{Fev07}. When operated in this manner, the system produces no DC-current, since every electron emission from the mesoscopic capacitor is followed by emission of a hole into the out-going edge state (when an electron is absorbed from the in-going edge state by the mesoscopic capacitor, a hole is in turn created in the out-going edge state). The mesoscopic capacitor also does not generate any zero-frequency noise. However, the mesoscopic capacitor produces an AC-current whose first Fourier component is $I_{AC}=2e/T$. Here, the factor of $2$ accounts for the two charges (one electron and one hole) that ideally are emitted during each period of the driving. Also, the noise power spectrum of the mesoscopic capacitor is generally different from zero at finite frequencies and, as we shall see, the finite-frequency noise provides a very useful characterization of the transport process.

\section{Theoretical Model}
\label{sec:model}

The system can accurately be described using scattering matrix theory expressed within a Floquet formalism which explicitly exploits the periodicity of the problem \cite{Mos02,Mos08}. In this approach the system is treated as being fully phase coherent and all quantum effects are appropriately accounted for. However, as well shall see, a simple semi-classical description of the mesoscopic capacitor is equally capable of explaining the measured current and noise properties. This model was first suggested by Mah\'e \emph{et al.} and shown to agree well with the measured data in numerical simulations of the experiment \cite{Mah10}. Here we analyze in more detail the model and find analytically the average current, the noise power spectrum, and the counting statistics of emitted electrons during a large number of periods.

\begin{figure*}
  \begin{center}
    \includegraphics[width=0.8\textwidth]{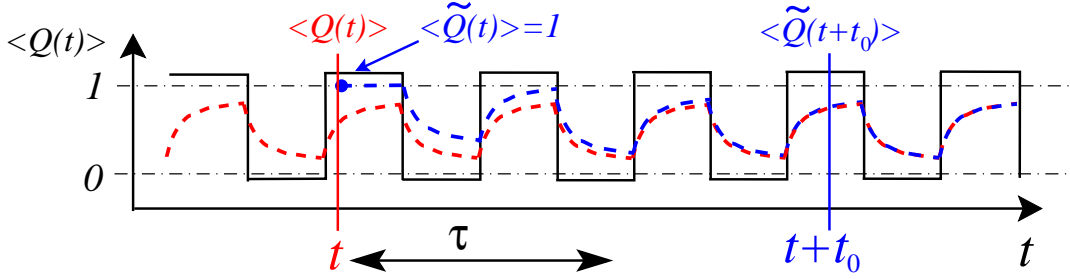}
    \caption{Schematics of periodic driving and (conditional) charge occupations. The mean occupation of the mesoscopic capacitor $\langle Q(t)\rangle$ is shown in red. The mean occupation is equal to the probability of the mesoscopic capacitor to be occupied with one (additional) electron. The blue curve shows the \emph{conditional} mean occupation $\langle \widetilde Q(t+t_0)\rangle$. This is the probability that the mesoscopic capacitor is charged with an (additional) electron at time $t+t_0$, given that it was charged at time $t$, such that $\langle \widetilde Q(t)\rangle=1$. Correlations decay on a time scale set by the correlation time $\tau$, implying that $\langle \widetilde{Q}(t+t_0)\rangle\simeq  \langle Q(t+t_0)\rangle$ for $t_0\gg \tau$.}
    \label{fig:calcnoise}
  \end{center}
\end{figure*}

The model can be explained by considering Fig.\ \ref{fig:system}a showing the time-dependent gate voltage $V_g(t)$. We discretize time in units of $\tau_o$ and make the following assumptions:  In the absorbtion phase, denoted by \textcircled{\footnotesize{1}}, the probability in each time step for an electron to enter the mesoscopic capacitor is $p$. With the amplitude of the periodic driving being on the order of the level spacing, higher-lying states can safely be neglected and only a single electron can enter the mesoscopic capacitor. Similarly, in the emission phase, denoted by \textcircled{\footnotesize{2}},  the probability of emitting the electron in each time step is $p$. This semi-classical picture can be formulated as a master equation in discrete time for the probability of the mesoscopic capacitor to be occupied by an electron. Setting the electron charge $e=1$ in the following, this probability is equal to the average (additional) charge of the mesoscopic capacitor $\langle Q\rangle$, where $Q=0,1$. The master equation determines the evolution of the average charge after one time step and reads \cite{Mah10}
\begin{equation}\label{master_eq}
  \langle Q(t_{k+1,\ell})\rangle= \left\{\begin{array}{ll}
     p[1\!-\!\langle Q(t_{k,\ell})\rangle]\!+\!\langle Q(t_{k,\ell})\rangle & \textcircled{\footnotesize{1}} \\
      \\
     (1-p)\langle Q(t_{k,\ell})\rangle & \textcircled{\footnotesize{2}}
    \end{array}\right. ,
\end{equation}
where we have used that $1-\langle Q\rangle$ is the probability for the mesoscopic capacitor to be empty and $t=t_{k,\ell}$ denotes time at the $k$'th time step during the $\ell$'th period. The absorbtion (emission) phase $\textcircled{\footnotesize{1}}$ ($\textcircled{\footnotesize{2}}$) corresponds to $k=1,2\ldots,K$ ($K+1,K+2,\ldots,2K$), where $K$ is the number of time steps in the absorbtion and emission phases, respectively, each of duration $T/2$.  In the experiment, the inverse noise measurement frequency was roughly equal to the driving frequency and both were much longer than the round trip time,  $2\pi/\omega \simeq T \simeq 60\,\tau_o$ \cite{Mah10}. This, in turn, allows us eventually to consider the continuous-time limit of the model, where the step size $\tau_o$ becomes irrelevant and drops out of the problem. Interestingly, the physics of the system is then governed by a single dimensionless quantity only, namely the ratio of the period $T$ and the escape, or correlation, time $\tau$ [defined in Eq. (\ref{eq:tau}) below]. Finally, we note that we have not provided a rigorous derivation of the model here, but instead we rely on the good agreement with experiments that we find. An improved understanding of the origin of the model as well as further comparisons with fully quantum coherent scattering matrix calculations will be the subject of future research.

\section{Average charge and current}
\label{sec:current}

We first consider the average current running out of the mesoscopic capacitor. The current can be related to the average charge $\langle Q\rangle$ as $\langle I(t)\rangle\equiv -\langle \dot{Q}(t)\rangle\simeq[\langle Q(t)\rangle-\langle Q(t+\tau_o)\rangle]/\tau_o$. It is easy to solve the master equation (\ref{master_eq}) for the average charge $\langle Q\rangle$ and we then obtain an expression for the average current which is similar to that of an $RC$ circuit:
\begin{equation}\label{qt2}
  \langle Q(t_{k,\ell})\rangle= \left\{\begin{array}{ll}
      \displaystyle 1-\beta_\ell\, e^{-(t_{k,\ell}-\ell T)/\tau} & \textcircled{\footnotesize{1}}  \\
       \\
      \displaystyle \alpha_\ell\, e^{-(t_{k,\ell}-[\ell+\frac{1}{2}]T)/\tau} & \textcircled{\footnotesize{2}}
    \end{array}\right. .
\end{equation}
Here, we have defined $\alpha_\ell=1/(1+\varepsilon)+\theta \varepsilon^{2\ell}$ and $\beta_\ell=1/(1+\varepsilon)-\theta \varepsilon^{2\ell-1}$ with $\theta$ depending on the initial conditions at the time when the periodic driving is turned on. The correlation time
\begin{equation}
\tau\equiv\frac{\tau_o}{\ln[1/(1-p)]}
\label{eq:tau}
\end{equation}
determines the time scale over which the system loses memory about the initial conditions encoded in $\theta$ and $\langle Q\rangle$ becomes periodic in time.

Figure \ref{fig:avcurrent}b,c illustrates, for two different values of $\tau/T$, the excellent agreement between numerical simulations of the current based on the master equation (\ref{master_eq}) and our  analytic result (\ref{qt2}), including the initial transient behavior. The mean charge emitted during the emission phase is $\tanh\left(T/4\tau\right)$ as we elaborate on below [in connection with Eqs.\ (\ref{pnotw}) and (\ref{eq:cumulants})].

\section{Noise power spectrum}
\label{sec:spectrum}

A more detailed characterization of the mesoscopic capacitor is provided by the noise power spectrum which recently was measured in the experiment by Mah\'{e} \emph{et al.} \cite{Mah10}. In the time domain the current-current correlation function is defined as $C(t,t_0)=\langle \delta I(t)\delta I(t+t_0)\rangle$, where $\delta I(t)=I(t)-\langle I(t)\rangle$ is the deviation from the average value of the current $\langle I(t)\rangle$. Due to the periodic gate voltage modulations, the system is not translational invariant in time, and the correlation function therefore does not only depend on the time difference $t_0$, but also on the absolute time $t$. Experimentally, the correlation function is averaged over the absolute time $t$ and the time-average correlation function $C(t_0)=\overline{\langle \delta I(t)\delta I(t+t_0)\rangle}^{\,t}$ is considered \cite{Mah10}. The noise power spectrum is then the Fourier transform of $C(t_0)$,
\begin{equation}
  \mathcal P_I(\omega)=\int_{-\infty}^{+\infty}dt_0\, \overline{\langle \delta I(t)\delta I(t+t_0)\rangle}^{\,t} e^{i\omega t_0}\,,
\end{equation}
or $\mathcal P_I(\omega)= (2/\tau_o^2)[1-\cos(\omega\tau_o)]\mathcal P_Q(\omega)\simeq \omega^2\mathcal P_Q(\omega)$ in terms of the corresponding time-averaged charge correlation function $\mathcal P_Q(\omega)=\int_{-\infty}^{+\infty}dt_0\, \overline{\langle \delta Q(t)\delta Q(t+t_0)\rangle}^{\,t} e^{i\omega t_0}$. Here, we have assumed $\omega\tau_o\ll 1$ as in the experiment \cite{Mah10}.

We evaluate the charge correlation function by following the schematics in Fig.\ \ref{fig:calcnoise}. First we note that $\langle Q(t)Q(t+t_0)\rangle$ is the joint probability for the capacitor to be charged with one electron both at time $t$ and at time $t+t_0$. Using conditional probabilities we then write $\langle Q(t)Q(t+t_0)\rangle=\langle Q(t)\rangle \langle \widetilde Q(t+t_0)\rangle$, where $\langle \widetilde Q(t+t_0)\rangle$ is the probability that the capacitor is charged with one electron at time $t+t_0$ given that it is charged at time $t$. For $t_0>0$, the conditional probability $\langle \widetilde Q(t+t_0)\rangle$ can be found by propagating forward in time the condition $\langle \widetilde Q(t)\rangle=1$ using the master equation (\ref{master_eq}), see also Fig.\ \ref{fig:calcnoise}. Similar reasoning applies to the case  $t_0<0$. Finally, integrating over $t$, the time-averaged charge correlation function becomes $\overline{\langle\delta Q(t)\delta Q(t+t_0)\rangle}^{\,t}=\,\frac{\tau}{T}\,e^{-|t_0|/\tau}\tanh\left(\frac{T}{4\tau}\right)$, and we immediately obtain the noise power spectrum
\begin{equation}\label{pnotw}
  \mathcal P_I(\omega)=\frac{2}{T}\tanh\left(\frac{T}{4\tau}\right)\,\frac{\omega^2\tau^2}{1+\omega^2\tau^2}.
\end{equation}
Interestingly, the noise power spectrum is given by the average charge emitted during the emission phase $\tanh(T/4\tau)$ (the factor of 2 accounts for the additional contribution from the average charge absorbed in the absorbtion phase) multiplied with a Lorentzian frequency dependence, which corresponds to the exponential decay of correlations in the time domain with time constant $\tau$, Eq.\ (\ref{eq:tau}). Finally, the factor $\omega^2\tau^2/T$ ensures that the noise spectrum becomes white in the high-frequency limit, while the zero-frequency limit $\mathcal P_I(0)=0$ reflects that charge on the average does not accumulate on the capacitor once $\langle Q\rangle$ has become periodic in time.

\begin{figure}
  \begin{center}
    \includegraphics[width=0.95\linewidth]{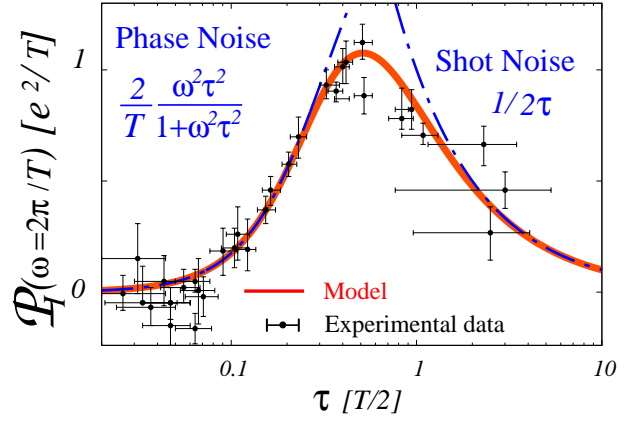}
    \caption{Noise power spectrum. Experimental results have been adapted from Ref.\ \cite{Mah10}. The full red line shows our analytic result, while blue dashed lines indicate limiting cases in the phase noise and shot noise regimes, respectively. The noise spectrum was measured at the frequency of the voltage modulations, $\omega=2\pi/T$,  as a function of the correlation time $\tau$.}
    \label{fig:noise}
  \end{center}
\end{figure}

Figure \ref{fig:noise} shows our analytic expression for the noise spectrum together with experimental results adapted from Ref.\ \cite{Mah10}. The analytic result captures the experimental data over the full range of correlation times $\tau$ and interpolates between the two limiting cases discussed by Mah\'{e} \emph{et al.} \cite{Mah10}. In the shot noise regime $\tau\gg T$, the probability of emitting and reabsorbing an electron during a single period is very small, and electron emission becomes rare. In this regime, we find $\mathcal P_I(\omega)\rightarrow 1/2\tau$ in agreement with Ref.\ \cite{Mah10}. In the phase noise regime $\tau\ll T$, the transport is highly regular and the probability of emitting and absorbing an electron during each period is close to one. In this case, the main source of finite-frequency fluctuations is the random times of emission and absorbtion within a period. In this limit, we find $\mathcal P_I(\omega)\rightarrow 2\omega^2\tau^2/[T(1+\omega^2\tau^2)]$ as proposed by Mah\'{e} \emph{et al.} \cite{Mah10}.

\section{Counting statistics}
\label{sec:fcs}

The noise power spectrum evaluated in the previous section concerns  the net current running out of the capacitor, including both absorbtion and emission of electrons. It is, however, interesting also to study the electron emission process alone. In this section, we thus consider the counting statistics of emitted electrons. To this end, we introduce the probabilities $P_j(n,\ell)$ that the capacitor is occupied by $j=0,1$ (additional) electrons, while $n$ electrons have been emitted after $\ell$ periods. The probability distribution for the number of emitted electrons is $P(n,\ell)=P_0(n,\ell)+P_1(n,\ell)$ and the corresponding cumulants $\llangle n^m\rrangle$ of the distribution are defined as $\llangle n^m\rrangle=\partial^m_{(i\chi)}\mathcal{S}(\chi,\ell)|_{\chi\rightarrow 0}$, where
\begin{equation}
\mathcal{S}(\chi,\ell)\equiv  \ln\left[\sum_n P(n,\ell)e^{in\chi}\right]
\end{equation}
is the cumulant generating function (CGF). The CGF can be written $\mathcal{S}(\chi,\ell)=\ln\left[\mathbf{1}\cdot \mathbf{P}(\chi,\ell)\right]$, where $\mathbf{1}=[1,1]^T$ and $\mathbf{P}(\chi,\ell)=[P_1(\chi,\ell),P_0(\chi,\ell)]^T$ with  $P_j(\chi,\ell)=\sum_n P_j(n,\ell)e^{in\chi}$, $j=0,1$. The evolution of the probability vector after one period of the driving is obtained from the master equation (\ref{master_eq}) and reads $\mathbf{P}(\chi,\ell+1)=\mathbf{A}(\chi)\, \mathbf{P}(\chi,\ell)$ with $\mathbf{A}(\chi)=\mathbf{L}_1^{s_1} \mathbf{L}_2^{s_2}(\chi)$ and $s_1+s_2=T/\tau_o$. The matrices
\begin{equation}
  \mathbf{L}_1=\left(\begin{array}{cc}1 & p\\ 0 & 1-p\end{array}\right)\;, \quad \mathbf{L}_2(\chi)=\left(\begin{array}{cc} 1-p & 0\\ p\,e^{i\chi} & 1 \end{array}\right)\,,
\end{equation}
generate one step of the evolution in the absorbtion ($\textcircled{\footnotesize{1}}$) and emission ($\textcircled{\footnotesize{2}}$) phases, respectively. The counting of emitted electrons is effected by the factor $e^{i\chi}$ entering the off-diagonal element of $\mathbf{L}_2(\chi)$ \cite{Bar03}.
Here, we have extended the model and allowed the absorbtion and emission phases to be of different durations. Only the total duration of the two phases has to equal a full period as expressed by the constraint $s_1+s_2=T/\tau_o$. For $s_1=s_2=T/2\tau_o$, we recover the original model \cite{Mah10,Alb10}.

The CGF after $\ell$ periods is now $\mathcal{S}(\chi,\ell)=\ln\left[\mathbf{1}\cdot\mathbf{A}^\ell(\chi)\mathbf{P}_{\mathrm{in}}\right]$ with $\mathbf{P}_{\mathrm{in}}$ being the $\chi$-independent initial condition as counting begins. For a large number of periods, $\mathbf{A}^\ell(\chi)$ is dominated by the largest eigenvalue $\lambda_0(\chi)$ of $\mathbf{A}(\chi)$, such that $\mathcal{S}(\chi,\ell)\rightarrow \ell \ln[\lambda_0(\chi)]$ \cite{Pis04}. The largest eigenvalue is
\begin{equation}
  \label{lambdaas}
  \lambda_0 (\chi)=f(\varepsilon_1,\varepsilon_2,\chi)+\sqrt{[f(\varepsilon_1,\varepsilon_2,\chi)]^2-\varepsilon_1 \varepsilon_2}
  \end{equation}
with
\begin{equation}
   f(\varepsilon_1,\varepsilon_2,\chi)=\frac{1}{2}\left[e^{i\chi}(1-\varepsilon_1-\varepsilon_2+\varepsilon_1\varepsilon_2)+\varepsilon_1+\varepsilon_2\right],
\end{equation}
and $\varepsilon_{1(2)}=(1-p)^{s_{1(2)}}$. For $s_1=s_2=T/2\tau_o$, we reproduce our result from Ref.\ \cite{Alb10}. With the CGF at hand we may now calculate the cumulants of the current $I= n/\ell$ after a large number of periods,  $\ell\gg 1$. However, before doing so, it is instructive to consider the CGF in different limiting cases.
\begin{figure}
  \begin{center}
    \includegraphics[width=0.45\textwidth]{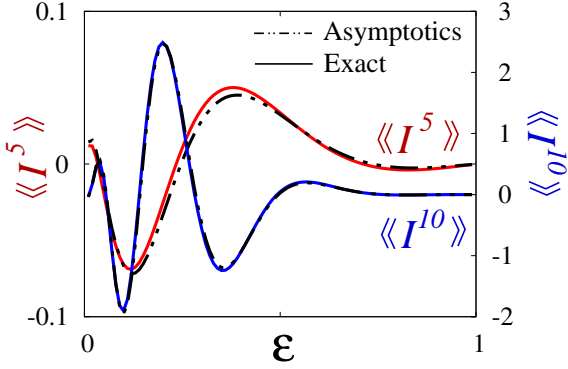}
    \caption{\label{fighighcumulant} High-order cumulants.  We show a comparison between exact results for the fifth and tenth cumultants, respectively, and the asymptotic approximation (\ref{eq:asympapprox}) as functions of the parameter $\varepsilon=e^{-T/2\tau}$. The approximation improves as the order is increased.}
  \end{center}
\end{figure}

In the shot noise regime $\tau\gg T$,  we may expand the CGF around $\varepsilon_{1}$ and $\varepsilon_{2}$ close to unity. We then find
\begin{equation}
\mathcal{S}(\chi,\ell)\simeq \ell\left[-\frac{\gamma_{1}+\gamma_{2}}{2}+\sqrt{\gamma_{1}\gamma_{2}e^{i\chi}+\left(\frac{\gamma_{1}-\gamma_{2}}{2}\right)^2}\right]
\end{equation}
corresponding to the CGF of a process where single electrons are absorbed at (the slow) rate $\gamma_1=1-\varepsilon_1\simeq 0$ and emitted at rate $\gamma_2=1-\varepsilon_2\simeq 0$. In the phase noise regime $\tau\ll T$, $\varepsilon_{1}$ and $\varepsilon_{2}$ are small, and an expansion of the CGF now yields
\begin{equation}
\mathcal{S}(\chi,\ell)\simeq \ell[i\chi+\varepsilon_1(e^{-i\chi}-1)+\varepsilon_2(e^{-i\chi}-1) ].
\label{eq:phasenoisecgf}
\end{equation}
Also this result has an immediate interpretation. The first term corresponds to a deterministic process in which exactly one electron is emitted in each cycle. The following two terms correspond to two independent Poisson processes describing ``cycle missing'' events occurring with rate $\varepsilon_1$ and $\varepsilon_2$ per period, respectively. Each of these processes create ``missing electrons'' in the otherwise regular stream of electrons that are emitted from the mesoscopic capacitor, either because it is not charged or because it fails to emit. Of course, if the mesoscopic capacitor is not charged it also fails to emit, however, such correlations only enter in higher orders of $\varepsilon_1$ and $\varepsilon_2$. Importantly, Equation (\ref{eq:phasenoisecgf}) allows us to estimate the accuracy of single electron emission in the phase noise regime. The failure rate is given by $\varepsilon_1+\varepsilon_2=(1-p)^{s_{1}}+(1-p)^{s_{2}}$ which under optimal conditions can be arbitrarily close to zero.

\section{Cumulants}
\label{sec:cumu}

From the CGF calculated in the previous section we can now calculate the cumulants of the electron emission process. For simplicity, we consider the case $\varepsilon_{1(2)}=(1-p)^{s_{1(2)}}=\varepsilon$. The first four cumulants of the current are then
\begin{equation}
\label{eq:cumulants}
\begin{split}
  \llangle I\rrangle &= \frac{1-\varepsilon}{1+\varepsilon}=\tanh\left(T/4\tau\right),\\
  \llangle I^2 \rrangle &= \frac{2\varepsilon}{(1+\varepsilon)^2} \llangle I\rrangle,\\
  \llangle I^3 \rrangle &= \frac{2\varepsilon(4\varepsilon-\varepsilon^2-1)}{(1+\varepsilon)^4}\llangle I\rrangle,\\
  \llangle I^4 \rrangle &= \frac{2\varepsilon(1-14\varepsilon+30\varepsilon^2-14\varepsilon^3+\varepsilon^4)}{(1+\varepsilon)^6}\llangle I\rrangle.\\
  \end{split}
\end{equation}
The first cumulant is the mean current of electrons, which equals the average charge emitted during the emission phase. The first cumulant varies from 0 in the shot noise regime ($\varepsilon\simeq 1$) to 1 in the phase noise regime ($\varepsilon\simeq 0$). In the shot noise regime the first few cumulants can be written as $\llangle I^m\rrangle\simeq(1/2)^{m-1}\llangle I\rrangle$, again indicating that the transport statistics is similar to that of a Poisson process with an effective charge of $1/2$. In the phase noise regime, the mean current is close to 1 and the first few cumulants are close to zero since electrons are emitted in an orderly manner due to the periodic driving. In this regime, low-frequency fluctuations are due to the rare cycle missing events.

The higher-order cumulants can be systematically  approximated by noting that the CGF  has square-root branch points at $i\chi_\pm=\ln[(1-\varepsilon)^2/4\varepsilon]\pm i\pi$, close to which the CGF behaves as $\mathcal{S}(\chi,\ell)\simeq 2\ell\sqrt{i\chi_\pm-i\chi}$. Following Ref.\ \cite{Fli09} we then find for large orders
\begin{equation}
\llangle I^m \rrangle \simeq \frac{4B_{m,-1/2}}{|i\chi_+|^{m-1/2}}\cos{\left[(m-1/2)\arg(i\chi_+)\right]},
\label{eq:asympapprox}
\end{equation}
where we have defined $B_{m,\mu}=\mu(\mu+1)\ldots(\mu+m-1)$. The asymptotic expression gives excellent agreement with exact results for the high-order cumulants as illustrated in Fig. \ref{fighighcumulant}.

\begin{figure}
  \begin{center}
    \includegraphics[width=0.45\textwidth]{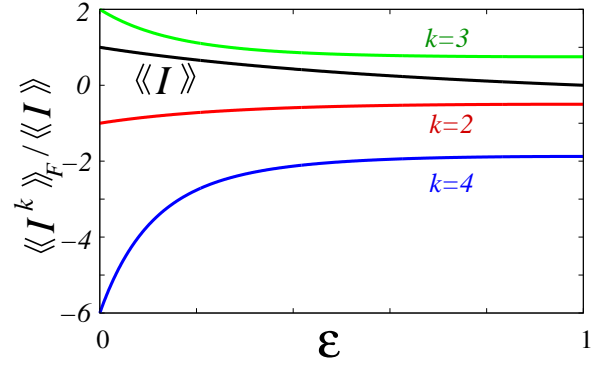}
    \caption{\label{figfaccumulant} Mean current and factorial cumulants. We show the mean current $\llangle I\rrangle$ together with the normalized factorial cumulants $\llangle I^k\rrangle_F/\llangle I\rrangle$, $k=2,3,4$, as functions of the parameter $\varepsilon=e^{-T/2\tau}$. As expected for non-interacting systems (see text), the factorial cumulants do not oscillate as functions of $\varepsilon$.}
  \end{center}
\end{figure}

Finally, we discuss the \emph{factorial} cumulants of the counting statistics. These are related to the ordinary cumulants as $\llangle I\rrangle_F=\llangle I\rrangle$,
$\llangle I^2\rrangle_F=\llangle I^2\rrangle-\llangle I\rrangle$, $\llangle I^3\rrangle_F= \llangle I^3\rrangle-3\llangle I^2\rrangle+2\llangle I\rrangle$, and in general, $\llangle I^m \rrangle_F =\sum_{j=1}^m s\left(m,j\right)\llangle I^j \rrangle$, where $s\left(m,j\right)$ are the Stirling numbers of the first kind. For the first few factorial cumulants we obtain
\begin{equation}
\label{eq:faccumulants}
\begin{split}
  \llangle I\rrangle_F &= \frac{1-\varepsilon}{1+\varepsilon}=\tanh\left(T/4\tau\right),\\
  \llangle I^2 \rrangle_F &= -\frac{1+\varepsilon^2}{(1+\varepsilon)^2} \llangle I\rrangle,\\
  \llangle I^3 \rrangle_F &= \frac{2(1+4\varepsilon^2+\varepsilon^4)}{(1+\varepsilon)^4}\llangle I\rrangle,\\
  \llangle I^4 \rrangle_F &= -\frac{6(1+9\varepsilon^2+9\varepsilon^4+\varepsilon^6)}{(1+\varepsilon)^6}\llangle I\rrangle\,.
  \end{split}
\end{equation}
Recently, it has been shown by Kambly \emph{et al.} that oscillations of the factorial cumulants as some parameter is varied must be due to interactions in the transport \cite{Kam11}. In contrast, for non-interacting systems the factorial cumulant do not oscillate and their over-all sign is determined only by the order. These statements are based on recent works by Abanov and Ivanov, who studied the zeros of the generating function for non-interacting electrons \cite{Aba08,Aba09}. Clearly, the factorial cumulants of the mesoscopic capacitor do not oscillate, Fig. \ref{figfaccumulant}. This is due to the non-interacting nature of the problem.

\section{Outlook}
\label{sec:outlook}

There are several open questions that deserve further investigation in future research. In the present work, we have described the mesoscopic capacitor using the model proposed by Mah\'{e} \emph{et al.} \cite{Mah10}. It is an important and interesting theoretical task to derive rigorously this model from a coherent scattering matrix description  \cite{Mos02,Mos08} and clearly identify the operating regimes in which it is valid. We have discussed the counting statistics of emitted electrons during a large number of periods in order to characterize fluctuations in the stream of charges. An alternative, and possibly more instructive, approach to this end would be to study the distribution of waiting times between subsequent emission events. This is a research direction that we are currently following. With the appearance of controllable single electron sources, a precise description and understanding of the emitted single particle states also become increasingly important \cite{Kee08}. Spectroscopy \cite{Mos11} and tomography protocols have recently been proposed \cite{Gre10}, and we believe that this line of investigation is likely to become a vivid field of research.

Also on the experimental side many interesting questions are still unaddressed. Currently, experiments are carried out with square-shape voltage modulations, unlike most theoretical descriptions for which sinusoidal voltage modulations are more convenient. It would be interesting to see experiments with harmonically varying gate potentials in order to better bridge theory and experiment. A measurement of the counting statistics could also provide further insights into the functionality of the device. Additionally, several applications of the mesoscopic capacitor have been put forward. These include proposals for the use of the mesoscopic capacitor as a quantum detector \cite{Fev08,Nig09} as well as for controlled two-particle effects \cite{Olk08,Spl09}.

\section{Conclusions}
\label{sec:conclu}

We have presented a short overview of the latest progress in understanding the fundamental noise properties of the mesoscopic capacitor. We have derived analytic expressions for the high-frequency noise spectrum which recently was measured. We extended the analysis to the counting statistics of emitted electrons after a large number of periods which we used to characterize the fluctuations in the stream of charges as well as to discuss the accuracy of the mesoscopic capacitor as a single electron source. Finally, we gave an outlook on possible directions for future theoretical and experimental research.

\section*{Acknowledgments}

We thank G.\ F\`eve, M.\ Moskalets and S.\ E.\ Nigg for useful discussions. The work was supported by the Swiss NSF and the NCCR Quantum Information and Technology.

\end{document}